\PassOptionsToPackage{sort&compress}{natbib}
\documentclass[num-refs]{wiley-article}

\usepackage{siunitx}
\newcommand{\TrtHz}{T/$\sqrt{\text{Hz}}$\ }

\usepackage{graphicx}
\usepackage{dcolumn}
\usepackage{bm}
\usepackage{xcolor}
\usepackage{soul}
\usepackage{hyperref}
\papertype{}

\title{Ultrawide Bandwidth Optomechanical Magnetometry Using Flux Concentration}

\author[1, 2, 3\authfn{1}]{Benjamin J. Carey}
\author[1,2,3\authfn{1}]{Nathaniel Bawden}
\author[1,2]{Fernando Gottardo}
\author[4,5]{James S. Bennett}
\author[6]{Douglas Bulla}
\author[6]{Scott Foster}
\author[1,2,3]{Warwick P. Bowen}

\contrib[\authfn{1}]{Equally contributing authors.}

\affil[1]{School of Mathematics and Physics, The University of Queensland, St Lucia, Queensland 4067, Australia.}
\affil[2]{ARC Centre of Excellence for Engineered Quantum Systems, St Lucia, Queensland 4067, Australia}
\affil[3]{ARC Centre of Excellence in Quantum Biotechnology, St. Lucia, Queensland 4067, Australia}
\affil[4]{Centre for Quantum Dynamics, Griffith University, Nathan, Queensland 4111, Australia.}
\affil[5]{School of Mathematical Sciences, Queensland University of Technology, Gardens Point, Queensland 4000, Australia.}
\affil[6]{Australian Government Department of Defence Science and Technology, Edinburgh, South Australia 5111, Australia}

\corraddress{Warwick P. Bowen, School of Mathematics and Physics, The University of Queensland, St Lucia, Queensland 4067, Australia.}
\corremail{w.bowen@uq.edu.au}

\fundinginfo{Defense Science and Technology Group (DSTG), Next Generation Technologies Fund (NGTF) and Advanced Strategic Capabilities Accelerator (ASCA); Australian Research Council Centre of Excellence for Engineered Quantum systems (EQUS,
 CE170100009): Australian Research Council Centre of Excellence in Quantum Biotechnology (QUBIC, CE230100021); Queensland Defense Science Alliance (QDSA); Orica Australia \textit{Pty Ltd}; Queensland Government, department of education, tourism, science, and innovation (Quantum 2032 Challenge)}

\runningauthor{Carey, Bawden et al.}

\begin{document}

\begin{frontmatter}
\maketitle

\begin{abstract}
Low-frequency magnetic fields carry vital information for neuroscience, navigation, and Earth science. However, they are generally weak, making it challenging to measure them with compact, room-temperature magnetometers. To overcome this challenge, we combine an on-chip optomechanical magnetometer with a high-permeability flux concentrator. Beyond boosting sensitivity and bandwidth, exploiting the concentrator’s nonlinear response converts low-frequency magnetic fluctuations into higher-frequency signals where the sensor is intrinsically most responsive. This sidesteps the technical noise that has long constrained the application of optomechanical magnetometry at low frequencies. Our measurements show order-of-magnitude improvements in sensitivity and extend performance into the sub-hertz regime, achieving $<$20\,n\TrtHz down to 3\,Hz and $<$100\,n\TrtHz at 0.1\,Hz. Because this approach requires no redesign of the underlying architecture, it can be readily applied across magnetometer technologies, opening the way to practical low-frequency sensing for applications from brain activity mapping to undersea navigation and biomedical diagnostics.

\end{abstract}
\end{frontmatter}

\section{\label{sec:introduction}Introduction}

Magnetic field sensing (magnetometry) enables technologies ranging from biomedical diagnostics \cite{brookes2022magnetoencephalography} to magnetic anomaly detection \cite{Sheinker2009}, GPS-free navigation and geophysical surveying \cite{Bennett2021}. Superconducting quantum interference devices (SQUIDs) \cite{stolz2022squids} or spin-exchange relaxation-free (SERF) magnetometers \cite{brookes2022magnetoencephalography} are the most sensitive technologies to date. However, the next generation of applications demands the development of chip-scale sensors that operate without cryogenic temperatures or magnetically shielded environments. Optomechanical magnetometers stand out in this regard, combining room-temperature, earth-field operation with the scalability of integrated photonics \cite{gotardo2023waveguide, Gotardo26}. Despite these potential advantages, their performance remains constrained by limited sensitivity and bandwidth, most notably at low frequencies where laser noise and other fluctuations become significant \cite{yu2016optomechanical}.

Optomechanical sensors exploit the dual enhancement of mechanical resonances, which provide enhanced signal response to stimuli, coupled to photonic resonances, which provide enhanced readout. This has enabled small-footprint devices for precision measurements of force \cite{harris2013minimum, fogliano2021ultrasensitive, melcher2014self}, acceleration \cite{bawden2025precision, Krause2012, chowdhury2023membrane}, sound \cite{basiri2019precision, wang2024high, McQueen2025}, pressure \cite{chen2022nano} and temperature \cite{liu2022ultrahigh}. They are particularly attractive for magnetic field sensing because of their compatibility with magnetostrictive materials, translating the intrinsically small magnetic-field-induced strains into mechanically amplified signals that can be read out optically with high sensitivity \cite{Forstner2014, gotardo2023waveguide, Li:18Quant, xu2024subpicotesla}. Recent demonstrations of on-chip devices have achieved sensitivities approaching 1\,p\TrtHz \cite{hu2024picotesla}, with larger devices achieving sensitives in the range of hundreds of femtoteslas 
\cite{xu2024subpicotesla, yang2024optomechanical}. However, their performance remains limited by 
the relatively small strain induced by magnetostrictive materials and by low-frequency noise. This not only restricts bandwidth and sensitivity but also obscures signals in the Hz--kHz frequency bands most relevant to neuroscience, navigation, and geophysics.

Addressing these limitations requires approaches that enhance sensitivity, extend bandwidth, and allow access to low-frequency signals. High-permeability flux concentrators provide one potential solution; by channeling magnetic flux into the sensing region, they can boost effective sensitivity without altering the underlying sensor architecture. Flux concentrators have been successfully employed to improve the performance of diamond \cite{fescenko2020diamond}, atomic vapor \cite{griffith2009miniature}, Hall effect \cite{leroy2008ac}, and SQUID-based \cite{matlashov2002high} magnetometers. However, they have not yet been applied with optomechanical magnetometers, nor have the been applied to facilitate low frequency magnetometery, a longstanding challenge for cavity optomechanical magnetometers \cite{basiri2019precision, yu2016optomechanical}. 


In this work, we demonstrate that millimeter-scale, 
high permeability structures can serve as flux concentrators to enhance the performance of silicon-chip-based optomechanical magnetometers, yielding an almost order-of-magnitude passive improvement in sensitivity, consistent with theoretical modeling. Further, we exploit the nonlinear magnetization of the concentrators to up-convert low-frequency magnetic signals into the resonance band of the mechanical sensor with very high conversion efficacy, 
retaining over 60\% of the original signal compared to less than 1\% efficacy from previous attempts \cite{Forstner2014}. This enables operation with flat sensitivity of better than 20\,n\TrtHz at frequencies down to 3\,Hz, and $<$100\,n\TrtHz down to 0.1\,Hz, significantly broadening the practical utility of on-chip optomechanical magnetometers for sub-Hz applications. Our approach could be readily applied to other optomechanical magnetometers which possess even better native sensitivity, such as those presented in \cite{hu2024picotesla}, potentially offering sensitivities beyond 10\,f\TrtHz at hertz or even sub-hertz frequencies. These flux concentrators could also be integrated with on-chip scalable production \cite{trindade2009control}. Together our results establish optomechanical sensing as a viable route to compact, room-temperature magnetometers capable of detecting low-frequency magnetic signals central to neuroscience, navigation, and Earth-science.

\section{Flux concentration}

\subsection{Magnetometer Design}
\label{sec:mag-design}
The optomechanical magnetometer used in this work is built on a freestanding silicon-on-insulator (SOI) platform with integrated optical readout. The fabrication process and operating principle are described in detail by Gottardo \textit{et al.} (2025) \cite{Gotardo26}, and a schematic of the device is shown in Fig.~\ref{fig:Experiment}(a) inset. 

The optical resonator comprises a 7-µm-long one-dimensional photonic cavity formed from a 600-nm-wide slotted waveguide terminated by a pair of photonic crystal mirrors. At one end of the cavity, a triangular cantilever (136\,µm base $\times$ 86\,µm height) serves as the mechanical transducer. A magnetostrictive galfenol film (Fe$_{82}$Ga$_{18}$, 600\,nm thick \cite{greenall2024QPM}) deposited on the cantilever converts applied magnetic fields into strain, driving mechanical motion. This motion perturbs the optical cavity resonance, producing a measurable frequency shift. 


Optical interrogation is performed by tuning a laser to the slope of the cavity resonance and detecting the reflected light. Magnetic-field-induced resonance shifts translate directly into intensity modulation of the reflected signal, enabling sensitive optical readout. Light is coupled on and off chip through an inverted taper waveguide, interfaced with a tapered optical fiber. To ensure robust operation, the fiber is permanently bonded to the waveguide facet using optical adhesive (NOA~86H) in a manner similar to that presented in \cite{wasserman2022cryogenic, Gotardo26}, eliminating the need for \textit{in-situ} alignment during measurement.

\subsection{Design of flux concentrators}
\label{subsec:FC-deign}
To effectively concentrate the magnetic field into the field-sensitive region of the optomechanical magnetometer we perform systematic modeling of the flux concentration system, both analytically and numerically. This allowed for the tailoring of concentrator geometry and material choice for maximum performance.

\subsubsection{Mathematical description of flux concentration}
\label{subsubsec:model}
For a high permeability ($\mu_r$) flux concentrator aligned with a uniform external magnetic field ($B_\text{ext}$), the internal magnetic flux density at its center is expressed as \cite{sun2013magnetic, leroy2006high} (see Sec.~S.1.A for detailed derivation)
\begin{equation}
    B_0 = \frac{\mu_r B_\text{ext}}{N(\mu_r - 1) + 1},
    \label{eq;B0-main}
\end{equation}
where $N$ is the demagnetization factor along the field axis, which is set by the geometry of the body \cite{leroy2006high} (see Sec.~S.1, Eq.~S3).  
For a flux-concentrating needle of length $l$, width $w$, and thickness $h$, 
the average field at the tip is well approximated by $B_0/m$, where $m = l / \sqrt{w \cdot h}$ is the needle aspect ratio and the axial field decays with distance $r$ from the tip as (see Sec.~S.1.A for details):
\begin{equation}
\label{eq:BR}
    B(r) = \frac{B_0}{m} + \left(B_\text{ext} - \frac{B_0}{m}\right)\cdot\left(\frac{l^2}{m^2r^2}+1\right)^{-\frac{1}{2}}.
\end{equation}

From this expression, the enhancement-factor (in dB) can be expressed 
as:
\begin{equation}
\label{eq:EF-B}
    \text{EF}(r) = 20 \log_{10} \left( \frac{B(r)}{B_\text{ext}} \right)
\end{equation}
These relations (Eqs.~\ref{eq;B0-main} \& \ref{eq:BR}) emphasize the two key parameters for the design of flux concentrators: magnetic permeability ($\mu_r$) and aspect ratio $m$. Increasing both of these will lead to enhanced concentration. This can be achieved with flux concentrator needles made of the high-$\mu_r$ material Metglas (see Sec.~S.1 and Fig.~S1).

\subsubsection{Numerical modeling of flux concentration}
\label{subsubsec:comsol}
To validate the analytical predictions and further inform the experimental flux concentrator design, we performed finite element method (FEM) simulations using COMSOL Multiphysics. A Metglas needle, 15\,mm long and 120\,µm $\times$ 25\,µm in cross-section, was placed in a uniform DC magnetic field oriented parallel to its length. The simulations, shown in Fig.~\ref{fig:Experiment}(b), clearly illustrate how the field lines are funneled toward the needle and concentrated near the tips, producing a strong local enhancement of the magnetic flux density.

Eq.~\ref{eq:EF} predicts a field enhancement of approximately 20\,dB at a sensor–-needle spacing of 30\,µm (averaged across the area of the magneto-mechanical transducer, as discussed in Sec.~\ref{sec:mag-design}), equivalent to an order-of-magnitude improvement in sensitivity. Moreover, the calculated decay of enhancement with increasing distance is in excellent agreement with the experimental results presented in Sec.~\ref{subsec:height}, providing confidence in both the model and the design choice of high-aspect-ratio Metglas microneedles.

To assess the frequency dependence of the magnetic field concentration from the concentrator, we also modeled its response to a uniform AC magnetic field of varying frequency and extracted the enhancement factor. The simulations reveal a flat enhancement spectrum up to approximately 100\,kHz, followed by a gradual roll-off at higher frequencies, with a reduction of approximately 2~dB at the mechanical resonance frequency of the optomechanical device (590~kHz) (see Sec.~S.4), confirming that the flux concentrator provides broadband performance across the frequency range accessible for optomechanical magnetometers.

\subsubsection{Fabrication of flux concentrators}
The modeling presented in Sec.~\ref{subsubsec:model}, \ \ref{subsubsec:comsol} \& S.1 predicts that flux concentration is maximized by combining high magnetic permeability with a large aspect ratio. To achieve this we fabricated microneedles from 25-µm-thick Metglas\textsuperscript{\textregistered} brazing foil (which boasts a relative permeability of $\mu_r \approx 10^5$). Needles of 125\,µm width and varying lengths were cut using a diamond-blade dicing saw, yielding aspect ratios ($m = l/\sqrt{w h}$) of $\sim$270. 
The width was chosen to be commensurate with the dimensions of the magnetostriction transducer used in the optomechanical device (136\,µm $\times$ 86\,µm), ensuring efficient spatial overlap between the concentrated magnetic field and the magnetostrictive element. 

Because the resulting needles are mechanically fragile, they were mounted onto $\sim$1\,mm-thick silicon chiplets for support. This provided rigidity for handling and precise alignment while leaving sufficient overhang to precisely position the tips close to the device during experiments. This mounting does not perturb the local magnetic environment, as Si has a negligible magnetic susceptibility.

\subsection{Experimental setup}
\begin{figure*}[htbp]
    \centering
    \includegraphics[width=\textwidth]{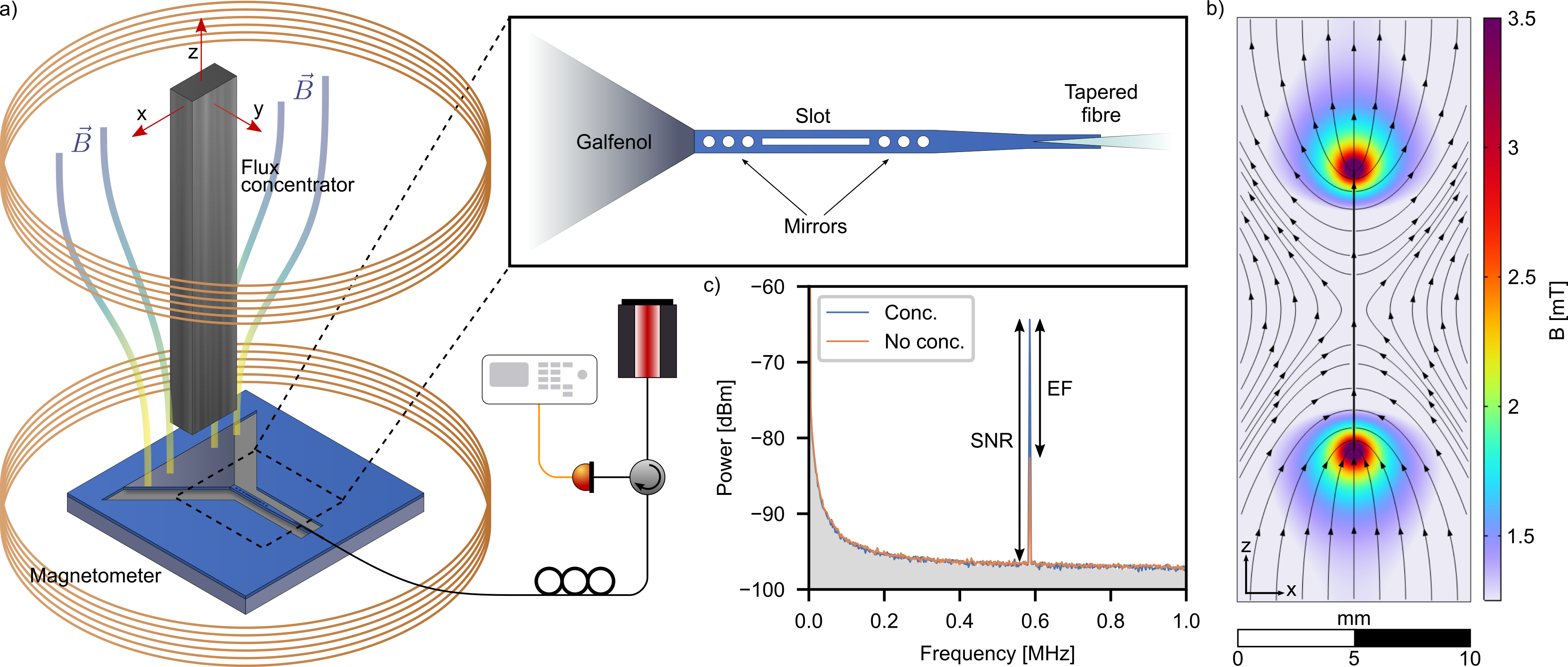}
    \caption{a) Schematic of the experimental setup. A Helmholtz coil produces a magnetic field that is directed onto the sensor by the flux concentrator. Inset: Device architecture. b) 2D cross-section of COMSOL simulation, demonstrating the magnetic field lines and relative flux density. c) Measured power spectra of a 590\,kHz 21\,\textmu{}T applied field with (conc.) and without (no conc.) the flux concentrator, with  signal-to-noise ratio (SNR) and enhancement factor (EF) emphasized. Here the resolution bandwidth (RBW) is 1.97\,kHz. The noise floor is indicated by the gray shaded region.}
    \label{fig:Experiment}
\end{figure*}

Magnetic field sensing experiments were carried out with the magnetometer positioned inside a Helmholtz coil, as shown in Fig.~\ref{fig:Experiment}(a), providing a uniform and well-calibrated field environment. A sinusoidal drive voltage was applied to the coil to generate magnetic fields with an RMS amplitude of 21\,µT, verified using a commercial teslameter (Lake Shore F71) and in agreement with theoretical calculations. The flux concentrator was mounted on a motorized three-axis translation stage for precision alignment above the magnetometer. This allowed reproducible control of the spacing between the needle tip and the magnetostrictive transducer, as informed by the modeling in Sec.~\ref{subsec:FC-deign}. The optical readout functions by monitoring shifts of the optical cavity resonance with a tunable external cavity diode laser (New Focus TLB-6730-P) which was tuned to the side of the resonance.

\subsection{Sensitivity enhancement}
To quantify the improvement in performance with the flux concentrator, the needle was positioned approximately 10\,µm above the center of the cantilever. An AC magnetic field was applied at 590\,kHz, corresponding to the fundamental cantilever resonance where the optomechanical sensor achieves peak magnetic sensitivity \cite{Gotardo26}. The result was a power spectrum with a sharp peak at the applied drive frequency, as depicted in Fig.~\ref{fig:Experiment}(c). The sensitivity $B_\text{min}$ at resonance was extracted from the known applied field $B_\text{app}$ and the measured signal-to-noise ratio (SNR in linear scaling) according to \cite{Forstner2014}:
\begin{equation}
    B_\text{min} = \frac{B_\text{app}}{\sqrt{\text{SNR} \cdot \text{RBW}}},
\end{equation}
where RBW is the resolution bandwidth of the spectrum analyzer.

With the concentrator in place, the device exhibited a peak sensitivity of 11\,n\TrtHz. This value is comparable to the best reported performance of integrated SOI magnetometers operated in vacuum, but here it is achieved in air, where squeeze-film damping typically degrades the mechanical response \cite{Gotardo26}. The enhancement of the signal power, in the linear response regime can be quantified through the enhancement factor:
\begin{align}
\label{eq:EF}
    \text{EF} &= 10 \log_{10}\left( \frac{P_\text{conc}}{P_\text{0}} \right) \ [\text{dB}],
\end{align}
where $P_0$ \& $P_\text{conc}$ are the measured signal power with and without flux concentration respectively. The enhancement factor improves the effective sensitivity (in-line with Eq.~\ref{eq:EF-B}) of the magnetometer when compared with the sensitivity when the concentrator is retracted as $B_\text{min}^\text{conc} = B_\text{min}/10^{\text{EF}/20}$. We found experimentally that the flux concentrator improved the sensitivity of the magnetometer from 89\,n\TrtHz to 11\,n\TrtHz, corresponding to an EF of 18\,dB, consistent with the order-of-magnitude B-field gain predicted in Sec.~\ref{subsec:FC-deign}~\&~S.1.

\subsection{Relative position}
\label{subsec:position}

\begin{figure}[h] 
    \centering 
    \includegraphics{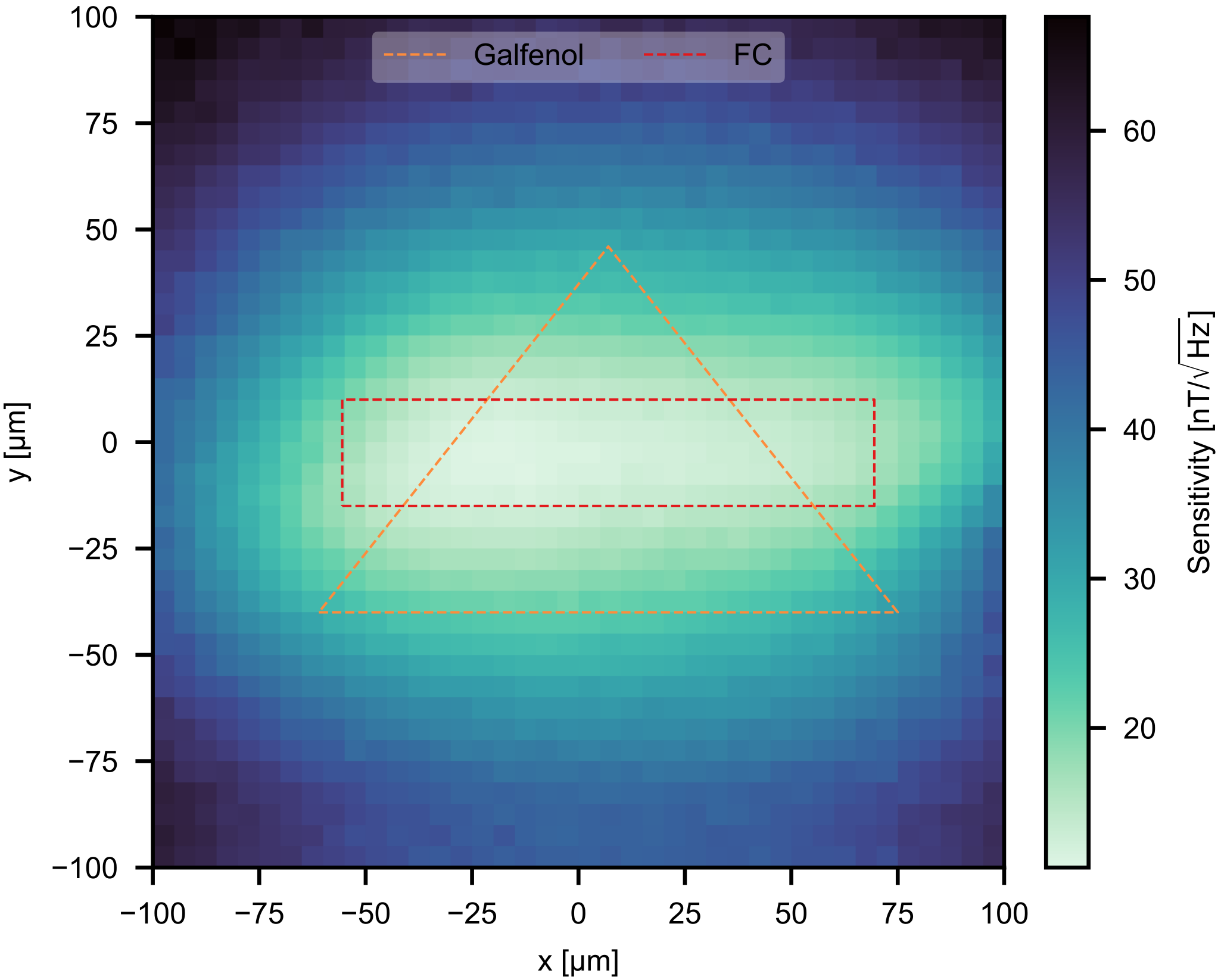} 
    \caption{Heat-map of the magnetometer sensitivity as the position of the flux concentrator is varied. The size of the galfenol and flux concentrator are indicated by the dashed lines.} 
    \label{fig:Map}
\end{figure}

To determine how the sensitivity depends on the transverse placement of the flux concentrator, 
a magnetic field was applied at the mechanical resonance frequency while the flux concentrator was scanned laterally across the sensor area at a fixed height of approximately 25\,\textmu{}m. 
The resulting sensitivity heat map, shown in Fig.~\ref{fig:Map}, reveals a localized enhancement when the concentrator is aligned with the cantilever, consistent with the field funneling predicted by the FEM simulations in Sec.~\ref{subsec:height} and depicted in Fig.~\ref{fig:Experiment}(b). The triangular shape of the cantilever is not fully resolved in the map because the lateral dimensions of the flux concentrator (136\,\textmu m) are comparable to the base of the triangular cantilever (125\,\textmu m). This introduces spatial averaging that smears out the cantilever profile. 

These results highlight the importance of geometric matching between the concentrated flux and the sensing element, wherein the external field sensitivity is optimized by maximizing the magnetic flux through the magnetostrictive cantilever.
This offers a potential pathway to even further improved performance in future devices via the use of tapered geometries (\textit{e.g.} \cite{fescenko2020diamond}) to increase total flux in, and hence the force exerted by, the magnetostrictive transducer. 

\begin{figure*}[htbp]
    \centering
    \includegraphics[width=\textwidth]{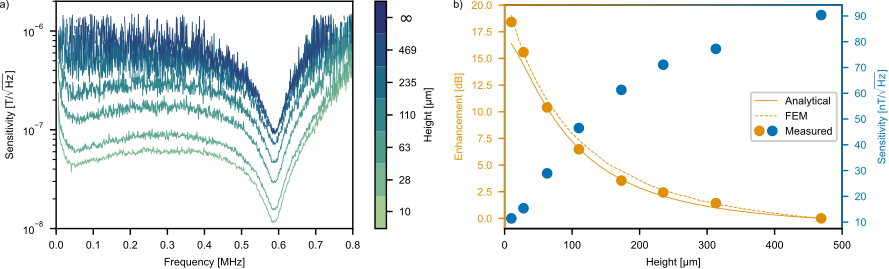}
    \caption{(a) Sensitivity spectra of the magnetometer for different flux concentrator heights above the device. (b) Measured and modeled enhancement of peak magnetic sensitivity as a function of flux concentrator height. Here, both the mathematical model, and FEM simulations results are determined using the geometry and permeability of the metglas flux concentrator.}
    \label{fig:HeightVariation}
\end{figure*}

\subsection{Height dependence and field decay}
\label{subsec:height}
To quantify the effect of vertical position and validate the modeling, the concentrator was positioned at the transverse location of maximum signal enhancement, and the sensitivity spectrum was recorded as the height of the flux concentrator above the device was varied. At each height, the frequency of the applied field was swept to measure the device sensitivity across a broad bandwidth, as shown in Fig.~\ref{fig:HeightVariation}. The sensitivity improvement is apparent across the entire measurement range and broadly frequency independent, with slightly more enhancement at lower frequencies ($<$200\,kHz), which arises from the frequency dependence of the complex permeability ($\mu = \mu'+i\mu''$) and is consistent with the frequency-dependent enhancement shown from the COMSOL modeling in Section~S.4.

The extracted peak sensitivities and corresponding enhancement factors (EFs) are summarized in Fig.~\ref{fig:HeightVariation}(b) and compared with predictions from both COMSOL simulations and the analytical model. 
The experimentally observed enhancement is in excellent agreement with both modeling approaches. The agreement between experiment and modeling validates the design approach and underscores the utility of these models for optimizing future flux concentrator geometries. 


\section{Operational bandwidth}
\label{sec-bandwidth}
In principle, the concentrator should provide enhancement across a wide frequency range, from towards DC where many application-relevant magnetic signals arise, up to a higher frequency roll-off in the 100s of kHz--MHz region, determined by the frequency dependence of the material's permeability (Sec.~S.4). This behavior is observed in Fig.~\ref{fig:HeightVariation}(a), with enhancement observed at all frequencies from 2~kHz to 800~kHz. A network analyzer was used to measure the sensitivity from 100\,Hz to 10\,kHz and from 5\,kHz to 1\,MHz. However, because of the resolution bandwidth of the spectrum analyzer used, the low-frequency sensitivity is not easily determined.

To evaluate the low frequency response, the sensitivity of our device was measured down to 1\,Hz in three overlapping frequency bands, as shown in Fig.~\ref{fig:MixUp}(a). Due to the resolution bandwidth (RBW) limitation of the analyzer, individual frequency response measurements were performed to assess the sensitivity at lower frequencies. The frequency ranges used in these measurements overlap, and the mutual agreement in the regions of overlap demonstrates the reliability and compatibility of the methods used.

The mechanical resonance peak is clearly observable at 590\,kHz, accompanied by a relatively flat sensitivity of $\sim$50\,n\TrtHz between 400\,kHz and 20\,kHz. Below this, the sensitivity degrades toward lower frequencies, proportional to $1/f^{0.7}$, caused by an increasing noise floor. This is consistent with previous reports of optomechanical sensors \cite{yu2016optomechanical, basiri2019precision}. To rule out laser noise as the source of this low-frequency degradation, the measurements were repeated using a significantly lower-noise, narrow-linewidth laser (NKT Photonics Adjustik E15), which yielded unchanged results. The sensitivity spectra were also measured with the flux concentrator removed, yielding the same low-frequency behavior, thus ruling out contributions of magnetic noise sources arising from within the concentrator, such as Barkhausen noise \cite{santa2019barkhausen}. Further measurements confirmed that electronic noise was also not responsible for the degradation. Therefore, the low frequency sensitivity degradation likely arises from a combination of environmental noise sources (\textit{e.g.} thermal, acoustic, vibrational) which manifest as phase or polarization fluctuations, or from fundamental noise sources within the device such as low-frequency stress noise within the cantilever.

\subsection{Nonlinear mixing based sensing}
\label{subsec:nonlinear}
To evade the low frequency noise, we implemented a nonlinear signal mix-up technique, similar in principle to the method demonstrated by Forstner \textit{et al.}~\cite{Forstner2014}, wherein a strong high-frequency local oscillator magnetic field is applied in order to drive the system into nonlinearity. Lower-frequency signals then manifest as sidebands around the local oscillator frequency. After detection, these sidebands can be demodulated to shift the signal back to its original frequency. Unlike the approach shown in \cite{Forstner2014}, where nonlinearity originated from the elasticity within the magnetostrictive element, here it arises from the magnetization nonlinearity of the flux concentrator (see Sec.~S.1.B).\\
This allows us to exploit the high magnetic susceptibility of Metglas to generate the mixed fields, while the galfenol remains in its optimal linear regime of magnetization and magnetostriction. This method has several distinct advantages. Firstly, it allows for operation at the mechanical resonance, and hence the maximum mechanical susceptibility, improving optical-noise-limited sensitivity (see Sec.~S.2). Secondly, because the nonlinearity exists outside of the magnetostrictive material, low-frequency thermomechanical noise is not mixed-up to the operational frequency. This is in direct contrast to Ref.~\cite{Forstner2014}, where the ultimate magnetic sensitivity is determined by the low-frequency thermomechanical strain noise in the magnetostrictive material, independent of the local oscillator field strength. Conversely, in our approach the magnetic field strength is independent of the low-frequency stress noise. 
Since the local oscillator amplifies the sideband generated by the low-frequency field without introducing noise, the sensitivity to low-frequency signals can in principle be enhanced beyond the on-resonance thermomechanical limit. This provides a pathway to overcome the fundamental noise source that has limited the sensitivity and bandwidth of optomechanical magnetometers to date.

The nonlinear response can be characterized by the second-order susceptibility (see Sec.~S.1.B for details),  
\begin{equation}
\label{eq:chi2}
    \chi^{(2)} = \frac{\mu_r^2}{M_{\text{S}}}      
    \left(\tanh^{3}{\left(\frac{ \mu_{r} H_{\text{DC}}}{M_{\text{S}}} \right)} -  
    \tanh{\left(\frac{ \mu_{r} H_{\text{DC}}}{M_{\text{S}}} \right)}\right),
\end{equation}
where $\mu_r$ is the relative permeability of the concentrator, $M_S$ its saturation magnetization, and $H_\text{DC}$ a biasing field used to tune the nonlinear response. A comparatively strong local oscillator (LO) field $B_\text{LO}$ is applied at the frequency of peak sensitivity (the mechanical resonance). A simultaneous low-frequency signal field $B_\text{sig}$ then mixes with $B_\text{LO}$ through the flux concentrator nonlinearity, producing sidebands at $\Omega_\text{LO} \pm \Omega_\text{sig}$:  
\begin{equation}
     B_\text{out}^{(2)} \approx \chi^{(2)}B_\text{sig} B_\text{LO} \cos\!\left[(\Omega_\text{LO} \pm \Omega_\text{sig})t\right].
\end{equation}
This effectively shifts the low-frequency signal into a higher-frequency band where noise is reduced, enabling detection well below the sensor’s baseband noise floor. The mix-up efficacy ($\eta^{(2)}$) and the predicted sensitivity of the mixed-up signal ($B_\text{min}^{(2)}$) are then given by (see Sec.~S.2):  
\begin{subequations}\label{eq:eta}
\label{eq:eta-Bmin}
\begin{align}
\eta^{(2)} = 2\chi^{(2)} B_\text{LO}
    = \frac{B_\text{out}^{(2)}}{Bsig}
    &= \frac{B_\text{min}(\Omega_0-\Omega)}{B_\text{min}^{(2)} (\Omega)}
    \label{eq:eta_a} \\
B_\text{min}^{(2)} (\Omega)
    &= \frac{B_\text{min}(\Omega_0-\Omega)}{\eta^{(2)}}.
    \label{eq:eta_b}
\end{align}
\end{subequations}
It can be seen, that increasing $\eta$ above unity would allow for the effective sensitivity to exceed that on the native unmixed signal. This is possible because the thermomechanical and optical noise (including shot-noise, thermo-refractive noise and laser phase noise) sources are constrained to the sensor, whilst the nonlinear mixing occurs within the flux concentrator, effectively isolating the mix-up process from these noise sources. There is potential for the emergence of noise originating with the concentrator itself, \textit{e.g.} Barkhausen noise, however this was not observed in our experiments.

\begin{figure*}[htbp]
    \centering
    \includegraphics[width=\textwidth]{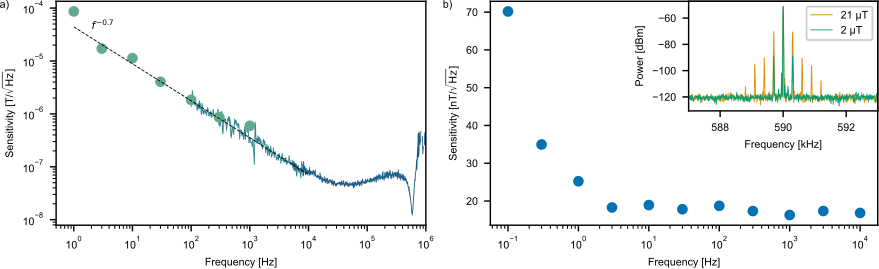}
    \caption{a) Wide frequency spectrum of the magnetometer sensitivity from 1\,MHz down to 1\,Hz. The flux concentrator was approximately 10\,\textmu m above the device and centered over the cantilever. The different colors represent different measurements. b) Low-frequency sensitivity using nonlinear mix-up. Inset: Power spectral density around the 590\,kHz mix-up frequency for 2\,\textmu T RMS and 21\,\textmu T RMS signal fields at 300\,Hz.}
    \label{fig:MixUp}
\end{figure*}

Equation~\ref{eq:eta-Bmin} reveals that the mix-up efficacy is determined only by $\chi^{(2)}$ and $B_\text{LO}$, with $\chi^{(2)}$ being limited by the material properties shown in Eq.~\ref{eq:chi2}; for metglas, it peaks at $\sim$60,000 (see Fig.~S2). In our experiment the DC bias was applied via a permanent NdFeB magnet held near the concentrator at a location that empirically optimized the second-order susceptibility, and hence the mix-up sensitivity. The practical limitation on $B_\text{LO}$ preventing an arbitrarily high mix-up efficacy arises from dynamic range limitations. This is demonstrated in Fig.~\ref{fig:MixUp}(b) inset; the nonlinear mixing between a 2-\textmu{}T (RMS) signal and a 40-\textmu{}T local oscillator produces a clean pair of sidebands for demodulation, whereas a 21-\textmu{}T signal produces multiple higher-order sidebands. This implies that $B_\text{LO}$ cannot be increased significantly beyond 40~\textmu T without signal degradation.  For our system ($\chi^{(2)} = 60,000 \ \& \ B_\text{LO} = 40$\,\textmu{}T), the expected mix-up efficacy is $\eta = 0.67$, predicting a slight trade-off in sensitivity for frequency conversion.

The sensitivity spectrum obtained from the mix-up experiment is presented in Fig.~\ref{fig:MixUp}(b). Here, an additional coil placed above the device generated the high-frequency drive at 590\,kHz with an RMS amplitude of $B_\text{LO} =$\,40\,\textmu{}T and a simultaneous low-frequency signal field of $B_\text{sig} =$\,2\,\textmu T RMS was applied, varied from 100\,mHz to 10\,kHz. 

We observe a sensitivity of $\sim$18\,nT/$\sqrt{\text{Hz}}$ for the nonlinearly mixed signal. This represents a mix-up efficacy of $\eta^{(2)} = 0.62$, in good agreement with the predicted value, and exceeding previous works exploiting nonlinearity in the magnetostrictive material by over two orders of magnitude \cite{Forstner2014}.
The sensitivity is consistent across the whole frequency band and agrees well with theoretical predictions (Fig.~S3), in contrast to Ref.~\cite{Forstner2014} which observed severe degradation in their mix-up efficacy above 1~kHz. Below 3~Hz the noise floor increases rapidly, degrading the sensitivity. The mix-up process evades the noise sources discussed earlier in Sec.~\ref{sec-bandwidth}, thus this degradation likely arises due to environmental magnetic field noise which also mixes up to the operational frequency as the measurement was not conducted in a magnetically shielded environment. Overall the mix-up performance spans nearly four decades of frequency and represents the first demonstration of optomechanical magnetometry at Hz to sub-Hz range frequencies without significant sensitivity degradation.

\section{\label{sec:discussion}Discussion and Conclusions}
This work establishes flux concentrators as a means to overcome the long-standing low-frequency performance limitations of chip-scale optomechanical magnetometers \cite{basiri2019precision, yu2016optomechanical}, whilst simultaneously improving sensitivity by an order of magnitude. We demonstrate an operational bandwidth of over seven orders of magnitude of frequency, extending below a hertz for the first time for an optomechanical magnetometer. The low frequency performance is achieved by exploiting the extremely high nonlinear mixing efficacy provided by high magnetic permeability materials, resulting in sensitivity close to the resonant thermomechanical noise floor even down to frequencies as low as 3~Hz. This enables the conversion of low-frequency magnetic signals to high-frequency sidebands with near-unity efficacy, compared to parts-per-thousand in previous works \cite{Forstner2014}. A further benefit of our approach lies in its versatility. Flux concentrators require no modification of the underlying sensor architecture, and thus, could be used to improve the sensitivity and bandwidth of state-of-the-art precision optomechanical magnetometers with minimal alterations. We anticipate that this will allow optomechanical magnetometers to excel in low-frequency applications that have previously been prohibitively challenging.
A key result from our work is that the use of a flux concentrator allows the evasion of thermomechanical noise. Thermomechanical noise has previously presented a fundamental limit to the performance of optomechanical magnetometers \cite{hu2024picotesla, Gotardo26}, and has been the dominant constraint on sensitivity on resonance in all previously reported experiments. Our experiments show a factor of eight improvement in on-resonance sensitivity beyond the thermomechanical noise floor. 
Moreover, while this is achieved by increasing the effective spatial footprint of the magnetometer -- drawing in flux from a larger area -- this does not come at the usual cost of increased thermomechanical noise. It is well known that the magnetostrictive response of optomechanical magnetometers can be improved by increasing the size of the magnetostrictive material ~\cite{xu2024subpicotesla, hu2024picotesla}. However, the thermomechanical noise level also increases with the square-root of the effective mass of the mechanical structure~\cite{hu2024picotesla, Forstner2012}, reducing the improvement in sensitivity. Employing a flux concentrator evades this constraint, increasing the effective size of the sensor without increasing its thermomechanical noise.\\

Beyond raw sensitivity improvements, and perhaps more significantly, the nonlinear mix-up technique demonstrated here provides a pathway to evade thermomechanical noise altogether. Previous mix-up techniques relied on the strain nonlinearity of the magnetostrictive material and were thus directly sensitive to thermomechanical strain noise \cite{Forstner2014}, even in the limit of a very large local oscillator field amplitude. By contrast, our technique evades this noise by employing a nonlinearity that is external to the optomechanical device. Increasing the magnitude of the oscillator field then allows the magnetic signal to be amplified above both the thermomechanical noise and the optical noise. For instance, in our experiments, the combination of this amplification and flux concentration allows a factor of 800 improvement in sensitivity for magnetic signals at 3\,Hz. The level of amplification will ultimately be limited by magnetic saturation within the flux concentrator leading to higher-order nonlinearities degrading the response. The sensitivity of the system may become ultimately limited by noise sources within the concentrator, such as Barkhausen noise. However, we observe no evidence of these effects in our experiments. The concentrators could, in principle, be implemented with planar geometries directly co-integrated on chip (\textit{e.g.}, \cite{fourneau2023microscale}). This makes them ideally suited for dense sensor arrays or magnetic assays, where localized enhancement and frequency up-conversion can be leveraged for high-resolution imaging and low-frequency detection.\\

The flux concentrator enhancement reported here opens practical pathways to apply optomechanical magnetometers to biomagnetism, neuroscience, navigation, geological surveying and magnetic anomaly detection which have proved challenging previously due to low-frequency noise. Beyond these immediate applications, combining planar flux concentrator integration with advances in on-chip photonic, and micro-mechanical designs could yield large-scale, low-power, high-sensitivity  sensor networks capable of addressing challenges where compact, broadband, and ultrasensitive magnetic field detection is required, including portable biomagnetic monitoring, magnetic surveying, and GPS-denied navigation.

\section*{Acknowledgements}
The authors acknowledge the facilities, and the scientific and technical assistance, of the Australian Microscopy $\&$ Microanalysis Research Facility at the Centre for Microscopy and Microanalysis, The University of Queensland. This work was performed in part at the Queensland node of the Australian National Fabrication Facility, a company established under the National Collaborative Research Infrastructure Strategy to provide nano- and microfabrication facilities for Australia’s researchers. The Commonwealth of Australia (represented by the Defence Science and Technology Group) supports this research through a Defence Science Partnerships agreement. This work was funded under the Next Generation Technology Fund and is being delivered through the Advanced Strategic Capability Accelerator. This work was also financially supported by the Australian Research Council Centre of Excellence for Engineered Quantum Systems (EQUS, Grant No. CE170100009), and the Australian Research Council Centre of Excellence in Quantum Biotechnology (QUBIC, Grant No. CE230100021). We thank the Queensland Defence Science Alliance (QDSA) for financial support of the project through the 2023 QDSA Collaborative Research Grant (CRG) funding round. This project was funded by the Queensland Government through the Department of Environment, Tourism, Science and Innovation’s (DETSI) Quantum 2032 Challenge Program. The program accelerates quantum sportstech, connects Queensland's research sector with industry, and showcases the state's quantum capabilities as part of Brisbane 2032's legacy. The authors acknowledge the highly valuable advice and support provided by Rodney Appleby and financial support by Orica Australia \textit{Pty Ltd}. The authors thank Professor James Macnae for his support and useful discussions of the research in this publication.


\section*{Supporting Information}

\printendnotes

\bibliography{References}

@article{Forstner2014,
author = {Forstner, Stefan and Sheridan, Eoin and Knittel, Joachim and Humphreys, Christopher L. and Brawley, George A. and Rubinsztein-Dunlop, Halina and Bowen, Warwick P.},
title = {Ultrasensitive Optomechanical Magnetometry},
journal = {Adv. Mater.},
volume = {26},
number = {36},
pages = {6348-6353},
year = {2014}
}

@ARTICLE{Sheinker2009,
  author={Sheinker, Arie and Frumkis, Lev and Ginzburg, Boris and Salomonski, Nizan and Kaplan, Ben-Zion},
  journal={IEEE Trans. Magn.}, 
  title={Magnetic Anomaly Detection Using a Three-Axis Magnetometer}, 
  year={2009},
  volume={45},
  number={1},
  pages={160-167},
  doi={10.1109/TMAG.2008.2006635}}

@Article{Bennett2021,
AUTHOR = {Bennett, James S. and Vyhnalek, Brian E. and Greenall, Hamish and Bridge, Elizabeth M. and Gotardo, Fernando and Forstner, Stefan and Harris, Glen I. and Miranda, Félix A. and Bowen, Warwick P.},
TITLE = {Precision Magnetometers for Aerospace Applications: A Review},
JOURNAL = {Sensors},
VOLUME = {21},
YEAR = {2021},
NUMBER = {16},
ARTICLE-NUMBER = {5568},
PubMedID = {34451010},
ISSN = {1424-8220},

}

@article{Forstner2012,
  title = {Cavity Optomechanical Magnetometer},
  author = {Forstner, S. and Prams, S. and Knittel, J. and van Ooijen, E. D. and Swaim, J. D. and Harris, G. I. and Szorkovszky, A. and Bowen, W. P. and Rubinsztein-Dunlop, H.},
  journal = {Phys. Rev. Lett.},
  volume = {108},
  issue = {12},
  pages = {120801},
  numpages = {5},
  year = {2012},
  month = {Mar},
  publisher = {American Physical Society},
  doi = {10.1103/PhysRevLett.108.120801},
  url = {https://link.aps.org/doi/10.1103/PhysRevLett.108.120801}
}

@article{Li:18Quant,
author = {Bei-Bei Li and Jan B\'{i}lek and Ulrich B. Hoff and Lars S. Madsen and Stefan Forstner and Varun Prakash and Clemens Sch\"{a}fermeier and Tobias Gehring and Warwick P. Bowen and Ulrik L. Andersen},
journal = {Optica},
number = {7},
pages = {850--856},
publisher = {Optica Publishing Group},
title = {Quantum enhanced optomechanical magnetometry},
volume = {5},
month = {Jul},
year = {2018},
url = {https://opg.optica.org/optica/abstract.cfm?URI=optica-5-7-850},
doi = {10.1364/OPTICA.5.000850},
}

@article{yu2016optomechanical,
  title={Optomechanical magnetometry with a macroscopic resonator},
  author={Yu, Changqiu and Janousek, Jiri and Sheridan, Eoin and McAuslan, David L and Rubinsztein-Dunlop, Halina and Lam, Ping Koy and Zhang, Yundong and Bowen, Warwick P},
  journal={Phys. Rev. Appl.},
  volume={5},
  number={4},
  pages={044007},
  year={2016},
  publisher={APS}
}

@article{Krause2012,
    author = {Krause, A. G. and Winger, M. and Blasius, T. D. and Qiang, L. and Oskar, P.},
    title = {A high-resolution microchip optomechanical accelerometer},
    journal = {Nat. Photon.},
    volume = {6},
    number = {11},
    year = {2012},
    pages = {768-772},
    issn = {1749-4885, 1749-4893},
    doi = {10.1038/nphoton.2012.245}
}

@article{gotardo2023waveguide,
  title={Waveguide-integrated chip-scale optomechanical magnetometer},
  author={Gotardo, Fernando and Carey, Benjamin J and Greenall, Hamish and Harris, Glen I and Romero, Erick and Bulla, Douglas and Bridge, Elizabeth M and Bennett, James S and Foster, Scott and Bowen, Warwick P},
  journal={Optics Express},
  volume={31},
  number={23},
  pages={37663--37672},
  year={2023},
  publisher={Optica Publishing Group}
}

@article{greenall2024QPM,
  title={Quantitative profilometric measurement of magnetostriction in thin-films},
  author={Greenall, Hamish and Carey, Benjamin J and Bulla, Douglas and Gotardo, Fernando and Harris, Glen I and Bennett, James S and Foster, Scott and Bowen, Warwick P},
  journal={Applied Surface Science},
  volume={662},
  pages={160105},
  year={2024},
  publisher={Elsevier}
}

@article{hu2024picotesla,
  title={Picotesla-sensitivity microcavity optomechanical magnetometry},
  author={Hu, Zhi-Gang and Gao, Yi-Meng and Liu, Jian-Fei and Yang, Hao and Wang, Min and Lei, Yuechen and Zhou, Xin and Li, Jincheng and Cao, Xuening and Liang, Jinjing and others},
  journal={Light: Science \& Applications},
  volume={13},
  number={1},
  pages={279},
  year={2024},
  publisher={Nature Publishing Group UK London}
}

@article{brookes2022magnetoencephalography,
  title={Magnetoencephalography with optically pumped magnetometers (OPM-MEG): the next generation of functional neuroimaging},
  author={Brookes, Matthew J and Leggett, James and Rea, Molly and Hill, Ryan M and Holmes, Niall and Boto, Elena and Bowtell, Richard},
  journal={Trends in Neurosciences},
  volume={45},
  number={8},
  pages={621--634},
  year={2022},
  publisher={Elsevier}
}

@article{stolz2022squids,
  title={SQUIDs for magnetic and electromagnetic methods in mineral exploration},
  author={Stolz, Ronny and Schiffler, Markus and Becken, Michael and Thiede, Anneke and Schneider, Michael and Chubak, Glenn and Marsden, Paul and Bergshjorth, Ana Bra{\~n}a and Schaefer, Markus and Terblanche, Ockert},
  journal={Mineral Economics},
  volume={35},
  number={3},
  pages={467--494},
  year={2022},
  publisher={Springer}
}

@article{basiri2019precision,
  title={Precision ultrasound sensing on a chip},
  author={Basiri-Esfahani, Sahar and Armin, Ardalan and Forstner, Stefan and Bowen, Warwick P},
  journal={Nature Communications},
  volume={10},
  number={1},
  pages={132},
  year={2019},
  publisher={Nature Publishing Group UK London}
}

@article{wang2024high,
  title={High resolution acoustic sensing based on microcavity optomechanical oscillator},
  author={Wang, Rong and Liu, WenYao and Pan, Ziwen and Fan, WenJie and Liu, Lai and Xing, Enbo and Zhou, Yanru and Tang, Jun and Liu, Jun},
  journal={Optics Express},
  volume={32},
  number={4},
  pages={4816--4826},
  year={2024},
  publisher={Optica Publishing Group}
}

@article{chowdhury2023membrane,
  title={Membrane-based optomechanical accelerometry},
  author={Chowdhury, Mitul Dey and Agrawal, Aman R and Wilson, Dalziel J},
  journal={Physical Review Applied},
  volume={19},
  number={2},
  pages={024011},
  year={2023},
  publisher={APS}
}

@article{fogliano2021ultrasensitive,
  title={Ultrasensitive nano-optomechanical force sensor operated at dilution temperatures},
  author={Fogliano, Francesco and Besga, Benjamin and Reigue, Antoine and Mercier de L{\'e}pinay, Laure and Heringlake, Philip and Gouriou, Clement and Eyraud, Eric and Wernsdorfer, Wolfgang and Pigeau, Benjamin and Arcizet, Olivier},
  journal={Nature Communications},
  volume={12},
  number={1},
  pages={4124},
  year={2021},
  publisher={Nature Publishing Group UK London}
}

@article{melcher2014self,
  title={A self-calibrating optomechanical force sensor with femtonewton resolution},
  author={Melcher, John and Stirling, Julian and Cervantes, Felipe Guzm{\'a}n and Pratt, Jon R and Shaw, Gordon A},
  journal={Applied Physics Letters},
  volume={105},
  number={23},
  year={2014},
  publisher={AIP Publishing}
}

@article{chen2022nano,
  title={Nano-optomechanical resonators for sensitive pressure sensing},
  author={Chen, Yanping and Liu, Shen and Hong, Guiqing and Zou, Mengqiang and Liu, Bonan and Luo, Junxian and Wang, Yiping},
  journal={ACS Applied Materials \& Interfaces},
  volume={14},
  number={34},
  pages={39211--39219},
  year={2022},
  publisher={ACS Publications}
}

@article{McQueen2025,
  title={Fibre-coupled photonic crystal hydrophone},
  author={McQueen, Lauren R and Bawden, Nathaniel and Carey, Benjamin J and Marinkovi{\'c}, Igor and Bowen, Warwick P and Harris, Glen I},
  journal={Optics Express},
  volume={33},
  number={12},
  pages={25910--25921},
  year={2025},
  publisher={Optica Publishing Group}
}

@article{xu2024subpicotesla,
  title={Subpicotesla Optomechanical Magnetometry},
  author={Xu, An-Ning and Li, Yifan and Li, Xiangliang and Liu, Bei and Liu, Yong-Chun},
  journal={Physical Review Letters},
  volume={133},
  number={15},
  pages={153601},
  year={2024},
  publisher={APS}
}

@article{harris2013minimum,
  title={Minimum requirements for feedback enhanced force sensing},
  author={Harris, Glen I and McAuslan, David L and Stace, Thomas M and Doherty, Andrew C and Bowen, Warwick P},
  journal={Physical review letters},
  volume={111},
  number={10},
  pages={103603},
  year={2013},
  publisher={APS}
}

@article{sun2013magnetic,
  title={Magnetic flux concentration at micrometer scale},
  author={Sun, Xu and Jiang, Lijun and Pong, Philip WT},
  journal={Microelectronic engineering},
  volume={111},
  pages={77--81},
  year={2013},
  publisher={Elsevier}
}

@article{leroy2006high,
  title={High magnetic field amplification for improving the sensitivity of Hall sensors},
  author={Leroy, Paul and Coillot, Christophe and Roux, Alain F and Chanteur, G{\'e}rard M},
  journal={IEEE Sensors Journal},
  volume={6},
  number={3},
  pages={707--713},
  year={2006},
  publisher={IEEE}
}

@article{fescenko2020diamond,
  title={Diamond magnetometer enhanced by ferrite flux concentrators},
  author={Fescenko, Ilja and Jarmola, Andrey and Savukov, Igor and Kehayias, Pauli and Smits, Janis and Damron, Joshua and Ristoff, Nathaniel and Mosavian, Nazanin and Acosta, Victor M},
  journal={Physical review research},
  volume={2},
  number={2},
  pages={023394},
  year={2020},
  publisher={APS}
}

@article{griffith2009miniature,
  title={Miniature atomic magnetometer integrated with flux concentrators},
  author={Griffith, W Clark and Jimenez-Martinez, Ricardo and Shah, Vishal and Knappe, Svenja and Kitching, John},
  journal={Applied Physics Letters},
  volume={94},
  number={2},
  year={2009},
  publisher={AIP Publishing}
}

@article{matlashov2002high,
  title={High sensitive magnetometers and gradiometers based on DC SQUIDs with flux focuser},
  author={Matlashov, AN and Koshelets, VP and Kalashnikov, PV and Zhuravlev, Yu E and Slobodchikov, V Yu and Kovtonyuk, SA and Filippenko, LV},
  journal={IEEE transactions on magnetics},
  volume={27},
  number={2},
  pages={2963--2966},
  year={2002},
  publisher={IEEE}
}

@article{leroy2008ac,
  title={An ac/dc magnetometer for space missions: Improvement of a Hall sensor by the magnetic flux concentration of the magnetic core of a searchcoil},
  author={Leroy, P and Coillot, C and Mosser, V and Roux, A and Chanteur, G},
  journal={Sensors and Actuators A: Physical},
  volume={142},
  number={2},
  pages={503--510},
  year={2008},
  publisher={Elsevier}
}

@article{fourneau2023microscale,
  title={Microscale Metasurfaces for On-Chip Magnetic Flux Concentration},
  author={Fourneau, Emile and Arregi, Jon Ander and Barrera, Aleix and Nguyen, Ngoc Duy and Bending, Simon and Sanchez, Alvaro and Uhl{\'\i}{\v{r}}, Vojt{\v{e}}ch and Palau, Anna and Silhanek, Alejandro V},
  journal={Advanced Materials Technologies},
  volume={8},
  number={16},
  pages={2300177},
  year={2023},
  publisher={Wiley Online Library}
}

@article{bawden2025precision,
  title = {Precision optomechanical accelerometer via hybrid test-mass integration},
  author = {Bawden, Nathaniel and Carey, Benjamin J. and Yeo, Poh-Meng and Arora, Nishta and Sementilli, Leo and Valenzuela, Victor M. and Romero, Erick and Harris, Glen I. and Wegener, Margaret and Bowen, Warwick P.},
  journal = {Phys. Rev. Appl.},
  volume = {24},
  issue = {6},
  pages = {064008},
  numpages = {9},
  year = {2025},
  month = {Dec},
  publisher = {American Physical Society},
  doi = {10.1103/knpw-1mdj}
}

@article{trindade2009control,
  title={Control of hysteretic behavior in flux concentrators},
  author={Trindade, IG and Leitao, DC and Pogorelov, Y and Sousa, JB and Chaves, RC and Cardoso, S and Freitas, PP},
  journal={Applied Physics Letters},
  volume={94},
  number={7},
  year={2009},
  publisher={AIP Publishing}
}

@article{liu2022ultrahigh,
  title={Ultrahigh-Resolution Optical Fiber Thermometer Based on Microcavity Opto-Mechanical Oscillation},
  author={Liu, Yize and Jiang, Junfeng and Liu, Kun and Wang, Shuang and Niu, Panpan and Xu, Tianhua and Zhang, Xuezhi and Wang, Ziyihui and Wang, Tong and Ding, Zhenyang and others},
  journal={Advanced Photonics Research},
  volume={3},
  number={9},
  pages={2200052},
  year={2022},
  publisher={Wiley Online Library}
}

@article{wasserman2022cryogenic,
  title={Cryogenic and hermetically sealed packaging of photonic chips for optomechanics},
  author={Wasserman, WW and Harrison, RA and Harris, GI and Sawadsky, Andreas and Sfendla, YL and Bowen, WP and Baker, CG},
  journal={Optics Express},
  volume={30},
  number={17},
  pages={30822--30831},
  year={2022},
  publisher={Optica Publishing Group}
}

@article{yang2024optomechanical,
  title={Optomechanical sensor network with fiber Bragg gratings},
  author={Yang, Shiwei and Zhang, Qiang and Yang, Linrun and Liu, Hanghua and Wang, Quansen and Zhang, Pengfei and Shen, Heng and Li, Yongmin},
  journal={arXiv preprint arXiv:2409.06943},
  year={2024}
}

@article{santa2019barkhausen,
  title={Barkhausen noise probes and modelling: A review},
  author={Santa-aho, Suvi and Laitinen, Arttu and Sorsa, Aki and Vippola, Minnamari},
  journal={Journal of Nondestructive Evaluation},
  volume={38},
  number={4},
  pages={94},
  year={2019},
  publisher={Springer}
}

@article{Gotardo26,
author = {Fernando Gotardo and Benjamin J. Carey and Nathaniel Bawden and Glen I. Harris and Hamish Greenall and Erick Romero and Douglas Bulla and James S. Bennett and Scott Foster and Warwick P. Bowen},
journal = {Optica},
number = {4},
pages = {558--565},
publisher = {Optica Publishing Group},
title = {Silicon-photonic optomechanical magnetometer},
volume = {13},
month = {Apr},
year = {2026}
}


\end{document}